\documentclass[twocolumn,showpacs,preprintnumbers,amsmath,amssymb]{revtex4}
\def\be{\begin{equation}}
\def\ee{\end{equation}}
\def\bea{\begin{eqnarray}}
\def\eea{\end{eqnarray}}
\def\ba{\begin{eqnarray*}}
\def\ea{\end{eqnarray*}}
\def\<{\langle}
\def\>{\rangle}
\def\~{\tilde}
\def\s{\sigma}

\def\b{\beta}

\def\o{\omega}
\def\t{\tau}

\newcommand{\Z}{\Bbb Z}

\newcommand{\av}[1]{\mbox{{\rm Av}}\left(#1\right)}

\newtheorem{theorem}{Theorem}

\begin{document}
\title{Local Order at Arbitrary Distances in Finite-Dimensional Spin-Glass Models}  
\author{Pierluigi Contucci}
 \email{contucci@dm.unibo.it}
\affiliation{Dipartimento di Matematica, Universit\`{a} di Bologna\\
             Piazza di Porta S.Donato 5, 40127 Bologna, Italy}
\author{Francesco Unguendoli}
 \email{unguendoli.francesco@unimo.it}
\affiliation{Dipartimento di Matematica Pura ed Applicata, \\
         Universit\`{a} di Modena
        e Reggio Emilia,\\ Via Campi 213/B, 41100 Modena, Italy}
\vskip .5truecm
\date{August 30th, 2004}
\begin{abstract}
For a finite dimensional spin-glass model we prove local order at low temperatures
for both local observables and for products of observables at arbitrary mutual distance.
When the Hamiltonian includes the Edwards-Anderson interaction we prove 
{\it bond} local order, when it includes the random-field
interaction we prove {\it site} local order. 
\end{abstract}
\maketitle

Spin-glass models in finite dimensions, in particular the nearest neighboor 
Edwards-Anderson and the random-field model, are among the most intensely studied 
models in condensed matter. Yet their low-temperature phase remains largely
unknown not only from a rigourous mathematical perspective but also from a
theoretical physics point of view (\cite{MPV,FH} and \cite{NS}). 

In this letter we show a simple rigorous proof of the local order property,
i.e. the fact that the overlap distribution concentrates close to one at 
low temperatures and at arbitrary distances between local observables. 

We consider models of Ising spin configurations
arranged in the d-dimensional lattice (for instance $\Z^d$); each spin interacts with a random esternal field
and with its nearest neighboors so that on a finite box $\Lambda$ the
Hamiltonian is
\be
H_\Lambda(\sigma) \; = \; - \sum_{i,j\in \Lambda \atop |i-j|=1}J_{i,j}\sigma_i\sigma_j - 
\sum_{i\in \Lambda} h_i\s_i\; ,
\label{hamil}
\ee
where the $J_{i,j}$ are i.i.d. Gaussian random variables with $\av{J_{i,j}}=0$ and
$\av{J^2_{i,j}}=\Delta^2$, the $h_i$ are i.i.d. Gaussian random variables with $\av{h_{i}}=0$ and
$\av{h^2_{i}}=\Gamma^2$. The $h$ and $J$ are moreover mutually indipendent: $\av{h_iJ_{k,l}}=0$
for all the $i,k,l$. The model of Hamiltonian (\ref{hamil}) reduces to the standard 
Gaussian Edwards-Anderson \cite{EA} when $\Gamma^2=0$ and to a pure random-field when 
$\Delta^2=0$. The (\ref{hamil}) is an important special case of the general Gaussian spin
glass model in which not only {\it sites} and nearest-neighboors {\it bonds} have
direct interactions but all the {\it subsets} $A$ of the lattice do have a polynomial 
interaction through $\s_A=\prod_{i\in A}\s_i$:
\be
H_\Lambda(\s) \; = \; -\sum_{A\subset \Lambda} J_A\s_A \; ,
\label{ghamil}
\ee
where all the $J$ are indipendent, centered, traslation invariantly distributed Gaussians.
The model of Hamiltonian (\ref{ghamil}) has been introduced in \cite{Ze} in the spirit of the 
general potentials of \cite{R} and subsequently studied in \cite{CG}. The (\ref{hamil}) 
is a special case of the (\ref{ghamil}) in which only the first two types of terms (sites and bonds)
give a contribution, but it also represents an approximation of it when the effect of the large
sets $A$ ($|A|\ge 3$) is simplified to a mean field one on each spin.

The local order is a concept introduced and developped within the classical
models without disorder in particular the ferromagnetic ones. 
For the Ising model it says that at low temperatures (large
$\beta$) the Boltzmann-Gibbs expectation  of the product of two sets of spins $\s_A=\prod_{i\in A}$ and
$\s_B=\prod_{j\in B}\s_j$ is close to $1$: 
\be\label{lro}
\omega (\s_A\s_B) \; \ge \; 1 - \frac{c}{\beta} \; ,
\ee
no matter how far the sets $A$ and $B$ are taken apart. In particular
since the constant $c$ is independent of both the volume of $\Lambda$ and
of the distance between the two sets the (\ref{lro}) remains true when
the thermodynamic limit is taken and successively the distance
between $A$ and $B$ is sent to infinity.

From such a property we know for instance that the equilibrium measure is concentrated on those 
configurations in which two spins are alligned, no matter how distant they are, and the lower the temperature
the sharper the concentration will be.

In a spin-glass the equilibrium state is described by the quenched measure i.e.
the Gaussian expectation of the random correlation functions. Due to simmetry
reasons quantities like $\omega (\s_i)$ or 
$\omega (\s_i\s_j)$ have zero Gaussian average so that the relevant physical
quantities are the average of their even powers:
$\av{\o^2(\s_i)}$, $\av{\o^2(\s_i\s_j)}$ etc.

Our main results are the following two theorems:
\begin{theorem}
The quenched state of Hamiltonian (\ref{hamil}) fulfills the site local order property 
at arbitrary distance 
when a random field is present $( \Gamma>0 )$:
for all the $m\in\Lambda$, $n\in\Lambda$ and independently of their distance
\be\label{slro}
\av{\o^2(\s_m\s_n)}  \; \ge \; 1 - \frac{s_1}{\b} - \frac{s_3}{\b^3} \; ,
\ee
where for all the $\Gamma>0$ 
\be
s_1=\frac{2}{\Gamma}\sqrt{\frac{2}{\pi}} \; ,\; \qquad
s_2=\frac{1}{\sqrt{2\pi}\Gamma^3}
\; .
\ee
\end{theorem} 
\begin{theorem}
The quenched state of Hamiltonian (\ref{hamil}) fulfills the bond local order property 
at arbitrary distance  when the 
two-body interaction is present $( \Delta>0 )$:
for all the $i,j\in\Lambda$, $|i-j|=1$, $k,l\in\Lambda$, $|k-l|=1$ and independently of the 
distance between the two bonds
\be\label{blro}
\av{\o^2(\s_{i,j}\s_{k,l})} \; \ge \; 1 - \frac{b_1}{\b} - \frac{b_3}{\b^3} \; , 
\ee
where for all the $\Delta>0$ 
\be
b_1=\frac{2}{\Delta}\sqrt{\frac{2}{\pi}} \; ,\; \qquad
b_2=\frac{1}{\sqrt{2\pi}\Delta^3} \; .
\ee
\end{theorem} 
It was soon realised \cite{MPV,G} that all 
the relevant quantities in the spin glass models are suitable expectations of the site-overlap
\be
q_i=\s_i\t_i \; ,
\ee
or the bond overlap
\be
q_{i,j}=\s_i\s_j\t_i\t_j \; .
\ee
Denoting by $\Omega$ the random Boltzmann-Gibbs
state over identical copies and defining the {\it quenched} measure by $<-> \, = \, \av{\Omega(-)}$
we have in fact
\bea
\av{\o^2(\s_i)}&=&\av{\o(\s_i)\o(\t_i)}=\\ \nonumber
&=&\av{\Omega(\s_i\t_i)} =\, <q_i> \; ,
\eea
and analogously
\be
\av{\o^2(\s_i\s_j)} = \, <q_{i,j}> \; .
\ee
According to this notation the (\ref{slro}) and (\ref{blro}) tell us that
the two quantities $<q_mq_n>$ and $<q_{i,j}q_{k,l}>$ are close to $1$ at low temperature
no matter how far the two spins (bonds) are taken.

The proof of the two theorems is computationally elementary and only uses the integration by parts formula
for centered Gaussian variables: let $\av{\zeta^2}=V_\zeta$ and $f$ a function of $\zeta$, then
\be\label{ibp}
\av{\zeta f(\zeta)} \; = \; V_\zeta\av{\frac{df}{d\zeta}} \; .
\ee
Let us carry out the proof for the general Hamiltonian
(\ref{ghamil}) and then we will apply it to the (\ref{hamil}). We introduce the convenient 
notation 
\be
\o_A \; = \; \frac{\sum_{\s} \s_A 
e^{-\b H_{\Lambda}(\s)}}{\sum_{\s} e^{-\b H_{\Lambda}(\s)}} \; 
\ee
and for every $J_A$ with $\av{J^2_A}=\Xi^2_A > 0$ we apply integration by parts to the quantity
$\av{J_A\o_A}$ 
\be\label{czz}
\av{J_A\o_A} \; = \; \Xi^2_A \beta (1-\av{\o^2_A}) \; .
\ee
Since an elementary estimate of the Gaussian integral gives (for $\Xi_A=\sqrt{\av{J^2_A}}$)
\be 
\av{J_A \o_A} \; \le \; \av{|J_A|} \; = \; \Xi_A\sqrt{\frac{2}{\pi}} \; ,
\ee 
we deduce from (\ref{czz}) that 
\be\label{note}
<q_A> \; = \; \av{\o^2_A} \; \ge \; 1 - \frac{1}{\Xi_A\beta}\sqrt{\frac{2}{\pi}} \; .
\ee
The previous formula says that the overlap distribuition of the set $A$ does
concentrates close to $1$ at low temperatures (see \cite{NS2}). 

To obtain the local order at arbitrary distances we proceed as follows: let consider
two subset $A$ and $B$ of $\Lambda$, and the notation:
\be\label{dot}
A \cdot B = A\cup B - A\cap B
\ee
we have
\be\label{kjh}
\s_A\s_B = \s_{A \cdot B}
\ee
because the sites $i$ in the intersections of $A$ and $B$ appear twice $\s_i^2=1$. Moreover
\bea
0&\le& \av{(\o_{A \cdot B}-\o_A\o_B)^2}=\\ \nonumber
&=&\av{\o^2_{A \cdot B}}+\av{\o^2_A\o^2_B} -2\av{\o_{A \cdot B}\o_A\o_B} \; ,\label{ldkfj}
\eea
so that
\be\label{wpw}
\av{\o^2_{A \cdot B}} \; \ge \; 2\av{\o_{A \cdot B}\o_A\o_B} - \av{\o^2_A\o^2_B} \; .
\ee
We apply then twice the (\ref{ibp}) to the positive quantity
\bea\label{asimm}
0&\le&\av{J^2_A(1-\o^2_B)}=\\ \nonumber
&=&\Xi^2_A\av{1-\o^2_B}-2\b^2\Xi^4_A\av{(\o_{A\cdot B}-\o_A\o_B)^2}\\ \nonumber
&-& 4 \beta^2\Xi^4_A \av{\o^2_{A} \o^2_{B}}
+4\beta^2\Xi^4_A \av{\o_{A\cdot B}\o_A\o_B}  \; ,
\eea 
from which 
\bea\label{owij} 
2\av{\o_{A\cdot B}\o_A\o_B}-\av{\o^2_{A} \o^2_{B}}\ge &&\\ \nonumber
\av{\o^2_{A} \o^2_{B}}+\av{(\o_{A\cdot B}-\o_A\o_B)^2}+&&\\ \nonumber
-\,\frac{1}{2\Xi^2_A\b^2}\av{1-\o^2_B} \ge && \\ \nonumber
\av{\o^2_{A} \o^2_{B}} -\frac{1}{2\Xi^2_A\b^2}\av{1-\o^2_B} \ge &&\\ \nonumber 
\av{\o^2_{A} \o^2_{B}}-\frac{1}{\sqrt{2\pi}\Xi^2_A\Xi_B\b^3}\; ,&&
\eea
where in the second inequality we have eliminated a positive term and in the
third we applyed the (\ref{note}).
Concatenating the (\ref{wpw}) to the (\ref{owij}) we get
\be\label{sega}
\av{\o^2_{A \cdot B}} \; \ge \;  \av{\o^2_{A} \o^2_{B}}-\frac{1}{\sqrt{2\pi}\Xi^2_A\Xi_B\b^3} \; .
\ee
We may then use the elementary inequality which states that for all
$a \le 1$, $b \le 1$
\be\label{ineq}
(1-a)(1-b) \; = \; ab-a-b+1 \; \ge \; 0 \; ; 
\ee
it implies 
\be\label{inq}
\av{\o^2_{A}\o^2_{B}} \; \ge \; \av{\o^2_{A}}+\av{\o^2_{B}}-1 \; ,
\ee
which together with (\ref{note}) gives
\be\label{inqi}
\av{\o^2_{A}\o^2_{B}} \; \ge \; 1-\left[ \frac{1}{\Xi_A} + \frac{1}{\Xi_B}
\right] \frac{1}{\beta}\sqrt{\frac{2}{\pi}} \; .
\ee
Putting together the (\ref{sega}) with (\ref{inqi}) we obtain
\bea\label{best}
&&\av{\o^2_{A\cdot B}} \; \ge  \\ \nonumber
&& 1 - \left[ \frac{1}{\Xi_A} + \frac{1}{\Xi_B}
\right] \frac{1}{\beta}\sqrt{\frac{2}{\pi}}
-\frac{1}{\sqrt{2\pi}\Xi^2_A\Xi_B}\frac{1}{\b^3}. 
\eea
The previous formula gives immediately Theorems 1 and 2 when applied to sites or to bonds.

It is interesting to observe that while for the ferromagnetic Ising model with zero 
magnetic field the property (\ref{lro}) is related to the phase transition of the model
and is called long range order (see \cite{R} for a general
definition and \cite{SML} for its proof in the 
two-dimensional Ising model), in the case of spin glasses our result is not
directly related to a phase transitions. A possible way to detect the existence
of a phase transition in a spin glass model would be in fact to bound from 
below the quantity $\av{\o^2(\s_m\s_n)}$ indipendently of $\Gamma$, which our theorem 1
fails to achieve. The method of integration by parts which we exploit here can
of course be iterated to two spins at distance $k$ but an easy computation show that
it leads to a bound $\av{\o^2(\s_1\s_k)}\ge 1 -\frac{c_k}{\beta}$ where the quantity
$c_k$ diverges for large $k$.
\begin{acknowledgments}
We thank M.Aizenman, A. Bovier, A. van Enter, C. Giardin\`a, S. Graffi, C. Newman, 
H. Nishimori, D. Stein and F.L.Toninelli for interesting discussions.
\end{acknowledgments}


\begin{thebibliography}{CU}
\bibitem{MPV} M.Mezard, G.Parisi, M.A.Virasoro, Spin Glass theory and
beyond, World Scien. (1987)
\bibitem{FH} D.S.Fisher and D.H.Huse, Phys. Rev. Lett., 56, 1601 (1986)
\bibitem{NS} C. M. Newmnan and D. L. Stein \\
http://arxiv.org/abs/cond-mat/0301403
\bibitem{EA} S.F.Edwards and P.W.Anderson, J.Phys.F. Vol. 5, May (1975)
\bibitem{Ze} B.Zegarlinski, Comm. Math. Phys, Vol 139, 305-339 (1991)
\bibitem{R} D. Ruelle, {Statistical Mechanics, Rigorous Results},
W.A.Benjamin, New York 1969
\bibitem{CG} P.Contucci, S.Graffi,
Jou. Stat. Phys., Vol. 115, Nos. 1/2, 581-589, (2004)
\bibitem{SML} T.D.Schultz, D.C.Mattis and E.H.Lieb, Rev.Mod.Phys., Vol. 36, 
856-871 (1964)
\bibitem{G} F.Guerra, Int. Jou. Mod. Phys. B, Vol. 10, 1675-1684, (1997)
\bibitem{NS2} C.M.Newmnan D.L.Stein, Phys.Rev. B, Vol 46, 973-982, (1992) 
\end{thebibliography}
\end{document}